\documentclass[twocolumn,footinbib,prb,superscriptaddress,floatfix]{revtex4-2}

\usepackage{amsmath,amssymb}
\usepackage{graphicx}
\usepackage{lipsum}
\usepackage{csquotes}
\usepackage{natbib}
\usepackage{amsthm}
\usepackage{mathtools}
\usepackage{physics}
\usepackage[T1]{fontenc}
\DeclareMathOperator{\sign}{sgn}
\usepackage{hyperref} 

\begin{document}
\title{Single-tone pulse sequences and robust two-tone shaped pulses for three silicon spin qubits with always-on exchange}

\author{David W. Kanaar}
    \email[Correspondence email address: ]{dkanaar1@umbc.edu}
    \affiliation{University of Maryland Baltimore County, Department of Physics, Baltimore, MD 21250, USA}
\author{Sidney Wolin}
\affiliation{University of Maryland Baltimore County, Department of Physics, Baltimore, MD, USA}
\author{Utkan G\"ung\"ord\"u}
\altaffiliation{Current address: Laboratory for Physical Sciences, College Park, Maryland 20740, USA}
\affiliation{University of Maryland Baltimore County, Department of Physics, Baltimore, MD, USA}
\author{J.~P.~Kestner}
\affiliation{University of Maryland Baltimore County, Department of Physics, Baltimore, MD, USA}

\date{\today} 

\begin{abstract}
Quantum computation requires high-fidelity single-qubit and two-qubit gates on a scalable platform. Silicon spin qubits are a promising platform toward realization of this goal. In this paper we show how to perform single-qubit and controlled-Z ({\sc CZ}) gates in a linear chain of three spin qubits with always-on exchange coupling, which is relevant for certain dot- and donor-based silicon devices. We also show how to make the {\sc CZ} gate robust against both charge noise and pulse length error using a two-tone pulse shaping method. The robust pulse maintains a fidelity of 99.99\% at 3.5\% fluctuations in exchange or pulse amplitude, which is an improvement over the uncorrected pulses where this fidelity can only be maintained for fluctuations in exchange up to 2\% or up to 0.2\% in amplitude.

\end{abstract}


\maketitle

\section{Introduction} \label{sec:Introduction}
Computational devices with quantum bits as their basis are predicted to have a wide range of applications such as breaking the widely deployed Rivest–Shamir–Adleman (RSA) encryption scheme \cite{Bernstein1998,NielsenandChuang}, molecular modeling \cite{Aspuru-Guzik2005,Peruzzo2014,Hempel2018}, finance \cite{Orus2019,Rebentrost2018,Bouland2020}, etc., although the practical extent of the desired quantum advantage remains to be seen. 
The main current challenge to exploring applications lies in making a quantum device which does not decohere before the desired computation is finished. Some current quantum devices contain sufficient number of qubits for specialized computations at the limits of what is presently achievable with classical computing \cite{Arute2019}, but not enough to make use of fault-tolerant error correcting codes, and they are too noisy to go beyond very shallow circuit depths without error correction. One possibility to enable greater circuit depths is to use dynamically corrected gates \cite{Khodjasteh2009,Jones2011,Calderon-Vargas2017,Edmunds2019}, i.e., control schemes designed such that the effects of coherent errors destructively interfere at the end of the evolution.

One promising candidate system for these quantum devices is spin qubits in silicon. Average fidelities for one-qubit gates in silicon have exceeded 99.9\% in Si/SiO$_2$ \cite{Yang2019} and Si/SiGe \cite{Yoneda2018} quantum dot devices with isotopically purified Si.
However, more than one-qubit gates are required for computation; a universal set of quantum gates is necessary, which can be obtained by adding an entangling two-qubit gate to the set of one-qubit rotations. Furthermore, in order to compose multi-qubit unitaries from this universal set, one must know how to do these one- and two-qubit gates without disturbing the other qubits.  This is not a trivial task when the interaction between qubits cannot easily be turned off, as is the case for silicon spin qubits in some dot
\cite{Veldhorst_2015,Watson_2018, Huang2019,chan2004} and donor \cite{Kalra_2014,Madzik_2020} systems.

In this paper we show how to perform a universal gate set in a three-qubit system in silicon with always-on exchange coupling.

Piecewise constant pulses have been implemented in two-qubit device experiments to perform entangling gates \cite{Watson_2018,Zajac_2018,Veldhorst_2015,Xue2019,Huang2019}. The reported two-qubit gate fidelities were between 78\% and 98\%, where many of the lower fidelities are limited mainly by charge noise. Theoretical high-fidelity two-qubit pulses have previously been proposed for an isolated pair of qubits \cite{Russ2018,Calderon-Vargas_2019}, and it has also been shown how to dynamically correct against charge noise \cite{Gungordu2020}. Uncorrected pulses have also been considered for a linear chain of three spin qubits \cite{Gullans_2019,chan2004,takeda_2020}. In this paper, we show how quasistatic charge noise can be corrected in a three-qubit system using a pulse shaping method from Refs.~\cite{Barnes2015,Gungordu2020}.
\par 

\section{Model} \label{sec:model}
\begin{figure}
    \centering
    \includegraphics[width=\linewidth]{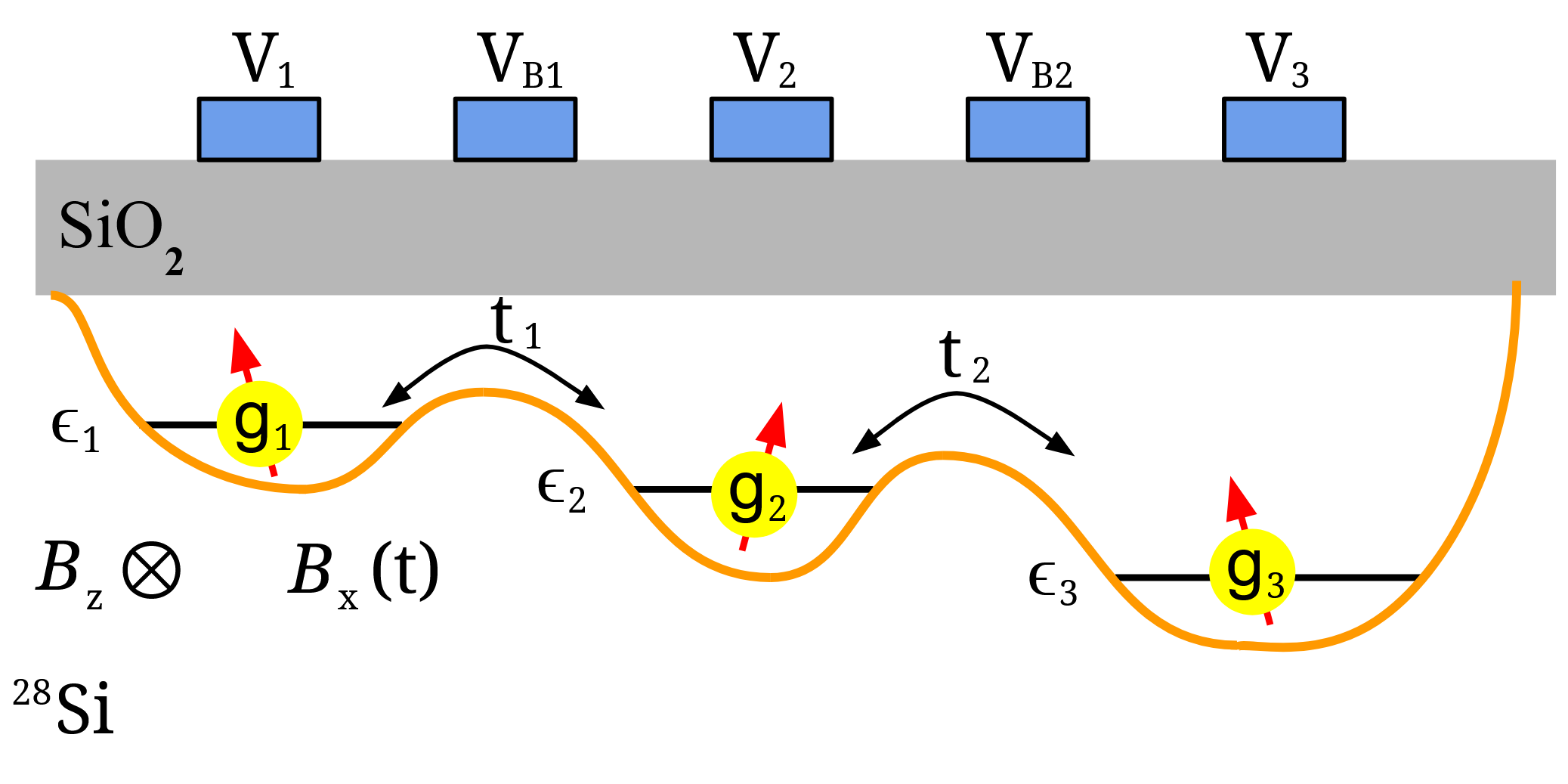}
    \caption{Illustration of a triple quantum dot within the two dimensional electron gas at the interface of Si/Si$\text{O}_2$ with metal gates on top. The gate voltages are tunable such that each dot is populated by one electron, and in some devices they can also be used to tune the effective $g$-factor of the electrons and the strength of the exchange couplings. The constant magnetic field in the z-direction, $B_z$, and the tunable magnetic field in the x-direction, $B_x(t)$, are shown as well.}
    \label{fig:1qubits}
\end{figure}

\newcommand{\bd}[0]{\textbf}
In this paper, we consider a three-qubit system comprising three quantum dots occupied by three electrons as shown in Fig.~\ref{fig:1qubits}, though our results also apply to exchange-coupled donor-based qubits. The occupation energies of each dot, $\epsilon_i$, differ due to the different applied voltages, $V_i$, on the corresponding top gates. There is a time-dependent magnetic control field, $\bd{B}_x (t)$, in the $x$-direction 
and a constant magnetic field, $\bd{B}_z$, in the $z$-direction. The tunneling energy $t_j$ between quantum dots $j$ and $j+1$ is tunable over a wide range via a barrier gate voltage in some setups \cite{Reed_2016,Zajac_2018,Yang_2020,Petit_2020,Leon_2020} but not in others \cite{Veldhorst_2015,Watson_2018, Huang2019}.

The low energy states of the system have one electron in each dot (in either the up or down spin state), and those states are coupled via virtual tunneling transitions. This means that, to second order in the tunneling, the effective Hamiltonian contains only coupling processes involving nearest neighbors of opposite spin virtually combining into either of the neighboring dots to form a singlet state. Higher order processes are negligible because the tunneling rate is typically small compared to the on-site Coulomb energy. The Hamiltonian can then be expressed in a Hubbard-style model as \cite{Yang2011}
   \begin{multline}
      H= \sum_{i=1}^3  \left(\frac{E_z^{(i)}}{2} \sigma_{Z}^{(i)} + \frac{E_{\perp}^{(i)}}{2} \sigma_{X}^{(i)}\right) \\ 
      + \sum_{i=1}^2\sum_{s=\uparrow,\downarrow} t_{i} \left(c_{i,s}^{\dagger}c_{i+1,s} + c_{i+1,s}^{\dagger}c_{i,s}\right)\\ 
      +\sum_{i=1}^3 \sum_{s=\uparrow,\downarrow} \epsilon_i n_{i,s} + \sum_{i=1}^3 U_i n_{i,\uparrow} n_{i,\downarrow},
   \end{multline}
where $E_{z,\perp}^{(i)}=\mu_b g_i B_{z,\perp}^{(i)}$ is the Zeeman energy of the electron in the $i$th quantum dot due to the magnetic field in the $z$($x$)-direction, $t_{i}$ is the tunneling energies to between opposite spin states of electrons in the  $i$th and $(i+1)$th quantum dot, $\epsilon_i$ is the on-site single electron occupancy energy of the $i$th quantum dot, $U_i$ is the on-site Coulomb energy associated with double occupation of the $i$th quantum dot, $\sigma_j^{(i)}$ is Pauli operator $\sigma_j$ on the electron in the $i$th dot, $c_{i,s}$ is the fermionic annihilation operator of an electron on the $i$th dot with spin $s$, and $n_{i,s} = c_{i,s}^{\dagger}c_{i,s}$ is the corresponding number operator. When using one electron spin resonance (ESR) driving field to drive all qubits, the ratios between the various $E_{\perp}^{(i)}$ are fixed, depending only on the $g$-factor differences and perhaps any difference in distances from each dot to the common ESR line, and the various $E_z^{(i)}$ need to differ from each other for single-qubit addressability. This is experimentally achievable by either having a magnetic field gradient or by manipulating the effective $g$ factors of the electrons \cite{Hanson2007}.   
Alternatively, in the case of electric dipole spin resonance (EDSR) driving, the various $E_{\perp}^{(i)}$ can be completely independent, naturally allowing single-qubit addressability. In the remainder of this work we will focus on the more restrictive ESR case, although our results are certainly applicable to the EDSR case as well.

In the experimentally relevant regime of $t_i,|E_{z}^{(i)}-E_{z}^{(i+1)}|  \ll U\pm(\epsilon_i-\epsilon_{i+1})$ for $i=1,2$, we can apply a Schrieffer-Wolff transformation to obtain the following effective spin Hamiltonian corresponding to low lying energy states
  \begin{equation}\label{eq:SWham}
      H =  \sum_{i=1}^3  \frac{E_z^{(i)}}{2} \sigma_{Z}^{(i)} + \frac{E_{\perp}^{(i)}}{2} \sigma_{X}^{(i)} +\sum_{i=1,2} J_i \vec{\sigma}^{(i)} \cdot \vec{\sigma}^{(i+1)},
  \end{equation}
up to $\mathcal O(t^2/U^2)$, where $J_i=\frac{2 t_i^2}{U_i+\epsilon_{i+1}-\epsilon_i}+\frac{2 t_i^2}{U_{i+1}+\epsilon_{i+1}-\epsilon_i}$ is the exchange coupling term between the $i$th and $(i+1)$th qubit. The exchange coupling can be controlled either via the detuning terms $\epsilon$ or, in devices equipped with a suitable barrier gate, via the tunneling rate. To reduce sensitivity to charge noise, it is preferable to operate at the symmetric operating point $\epsilon_i = 0$ \cite{Reed_2016}, but then one is left with a fixed, always-on residual exchange $\sim t^2/U$ in the absence of tunnel barrier control. (In fact, even when barrier control is possible and tunneling can be turned off completely, the magnetic dipole-dipole coupling should not be forgotten \cite{deSousa_2004} and provides a lower bound on $J$ of order 10 kHz for 30 nm qubit spacing. While small compared to MHz exchange coupling, this is not negligible when attempting to reach fidelities of 99.99\% and above.)
Our focus below is on the case of always-on exchange coupling.

We also note that, although for specificity our focus is on the case of dot-based systems, the same situation is quite relevant for weakly exchange-coupled electronic donor spins. In that case, one once again one has local ESR driving, single-qubit addressability through different Zeeman energies (either from initializing different nuclear spin states of the donors \cite{Kalra_2014} or through electrically adjusting the $g$-factors \cite{Hill_2005,Laucht_2015}), and an exchange coupling that is advantageous to leave fixed \cite{Kalra_2014,Madzik_2020}. 
  
To further simplify Eq.~\eqref{eq:SWham} we move to the rotating frame that eliminates the $E_z^{(i)}$ terms, which are typically several orders of magnitude larger than the other terms \cite{Huang2019}. 
In order to allow more than one qubit to be resonantly driven at the same time, we assume the possibility of using a two-tone driving field. The rotating frame Hamiltonian,  $H_R=R H R^{\dagger}+i \hbar (\partial_t R)  R^{\dagger}$, with two-tone driving can be written in terms of Pauli matrices as ignoring the identity terms which contribute an unimportant global phase
  \begin{multline} \label{eq:HR}
      H_R = \frac{J_1}{4} \sigma_{Z}^{(1)}\sigma_{Z}^{(2)}+ \frac{J_2}{4} \sigma_{Z}^{(2)} \sigma_{Z}^{(3)} \\
      +\sum_{j=1}^2\sum_{i=1}^3 \frac{\Omega_{j}^{(i)}}{2}\left(\cos{(( E_z^{(i)}-\omega_j)t+\phi_j)} \sigma_X^{(i)} \right. 
      \\
      +\left. \sin{(( E_z^{(i)}-\omega_j)t+\phi_j)} \sigma_Y^{(i)}\right)
  \end{multline}
where $E_{\perp}^{(i)}$ has been replaced with $\sum_{j=1}^2\Omega_j^{(i)} e^{i \left(\omega_j t + \phi_j\right)}$ as they are effectively equivalent using the rotating wave approximation, with $\Omega_j^{(i)}$ being the real time-dependent envelope of the $j$th tone of the oscillating driving field at the location of the $i$th qubit and $\phi_j$ the (generally, time-dependent) phase. In the experimental settings we are considering, $E_z^{(i)} \sim 10$GHz and $|E_z^{(i)}-E_z^{(i+1)}| \sim 10$MHz \cite{Huang2019}, much larger than any other energy scale in the Hamiltonian. So, we use the rotating wave approximation and ignore any $\Omega_j^{(i)}$ term for which the frequency $\omega_j$ of the driving tone is not resonant with the corresponding $2 E_z^{(i)}$. The phase $\phi_j$ of the driving field is also accurately controllable experimentally, so one can control the two single-qubit axes in Eq.~\eqref{eq:HR} independently.

There are also sources of noise in the Hamiltonian, and in general the largest source is charge noise, which causes stochastic shifts to the electrostatic triple-well potential of Fig.~\ref{fig:1qubits}. The main effect of charge noise is on the coupling term, $J_i \xrightarrow[]{} J_i+\delta J_i$. Charge noise can also cause fluctuations in the g-factor leading to Zeeman noise, however these fluctuations are on the order $10^{-9}$ so they will be neglected \cite{Ruskov2018,Veldhorst_2014,HuangP2014,Yoneda2018}. Noise in the control electronics can also result in driving amplitude noise such that $\Omega_j^{(i)} \xrightarrow[]{} \Omega_j^{(i)}(1+\delta \alpha_i)$ \cite{vanDijk2019}.

\section{Piecewise constant pulse sequences} \label{sec:pulses}
In this section we will show how to straightforwardly perform exact one- and two-qubit gates in the three-qubit system in the absence of noise. We also outline how this idea can be extended to noise-suppressing composite sequences as well.

An arbitrary local rotation can be Euler decomposed as 
\begin{equation}\label{eq:Euler1}
R = e^{-i \frac{\alpha}{2} \sigma_{Z}} e^{-i \frac{\beta}{2} \sigma_{X}} e^{-i \frac{\gamma}{2} \sigma_{Z}},
\end{equation}
or in terms of only $z$ rotations and a $\pm \pi/2$ rotation about $x$,
\begin{equation}\label{eq:Euler2}
R = e^{-i \frac{\alpha}{2} \sigma_{Z}} e^{-i \frac{\pi}{4} \sigma_{X}} e^{-i \frac{\beta}{2} \sigma_{Z}} e^{-i \frac{-\pi}{4} \sigma_{X}} e^{-i \frac{\gamma}{2} \sigma_{Z}}.
\end{equation}
In fact, one only needs either a $+\pi/2$ or a $-\pi/2$ $x$ rotation, since
\begin{align}\label{eq:Euler3}
R &= e^{-i \frac{\alpha}{2} \sigma_{Z}} e^{-i \frac{\pi}{4} \sigma_{X}} e^{-i \frac{\beta+\pi}{2} \sigma_{Z}} e^{-i \frac{\pi}{4} \sigma_{X}} e^{-i \frac{\gamma+\pi}{2} \sigma_{Z}}
\\
&= e^{-i \frac{\alpha+\pi}{2} \sigma_{Z}} e^{-i \frac{-\pi}{4} \sigma_{X}} e^{-i \frac{\beta+\pi}{2} \sigma_{Z}} e^{-i \frac{-\pi}{4} \sigma_{X}} e^{-i \frac{\gamma}{2} \sigma_{Z}}.
\end{align}

Thus, the ability to do a $\pm \pi/2$ $x$ rotation along with virtual $z$ rotations \cite{McKay2017} (which are instantaneous and error free) suffices to generate an arbitrary single-qubit rotation.  Adding a {\sc CZ} gate between nearest neighbors completes a universal gate set.

The main idea below is to decompose the Hamiltonian \eqref{eq:HR}, which lies in $\mathfrak{su}(8)$, into $\mathfrak{su}(2)$ and $\mathfrak{u}(1)$ subalgebras and then use Euler angle decomposition within each $\mathfrak{su}(2)$ such that the overall effect is to perform a desired operation on the intended qubit while returning the idle qubits to their original state despite an always-on interaction. In Sec.~\ref{sec:pulseresult} we will use a similar $\mathfrak{su}(2)$ approach to accomplish the same goal while also dynamically correcting errors by substituting pulse shaping theory for Euler angle decomposition. 

\subsection{\protect{$-\frac{\pi}{2}$ $x$} Rotation on an Outer Qubit}
Suppose we wish to rotate qubit 1. With single-tone driving at the resonant frequency of qubit 1, i.e., $\omega_1 = E_z^{(1)}$ and $\Omega_2^{(i)}=0$, and taking $\Omega_1^{(1)}=\Omega$ for simplicity of notation, the Hamiltonian \eqref{eq:HR} simplifies to one that lives in an $\mathfrak{su}\left(2\right)\oplus\mathfrak{u}\left(1\right)$ subalgebra, 
\begin{align}\label{eq:CZdecomp1}
    H &= \frac{J_1}{4} \sigma_{Z}^{(1)}\sigma_{Z}^{(2)} + \frac{\Omega}{2} \sigma_{X}^{(1)} +\frac{J_2}{4} \sigma_{Z}^{(2)}\sigma_{Z}^{(3)}
    \\
    &= H_{\mathfrak{u(1)}} + H_{\mathfrak{su(2)}}
\end{align}
where
\begin{align}\label{eq:CZdecomp2a}
    H_{\mathfrak{u(1)}}&= \frac{J_2}{4} \sigma_{Z}^{(2)}\sigma_{Z}^{(3)},
    \\ \label{eq:CZdecomp2b}
    H_{\mathfrak{su(2)}} &= \frac{J_1}{4} \sigma_{Z}^{(1)}\sigma_{Z}^{(2)} + \frac{\Omega}{2} \sigma_{X}^{(1)}.
\end{align}
The two generators in $H_{\mathfrak{su(2)}}$ have the same commutation properties as two Pauli operators while the $H_{\mathfrak{u(1)}}$ term commutes with both. So regardless of the driving, the evolution always contains a factor of $e^{\left(-i \frac{J_2 t}{4} \sigma_{Z}^{(2)}\sigma_{Z}^{(3)}\right)}$ and to avoid unwanted entanglement of qubits 2 and 3 during the rotation of qubit 1 the total pulse time must be an integer multiple of $2\pi/J_2$.
 
Consider first the evolution due to $H_{\mathfrak{su(2)}}$.  Note that the choices $\Omega = \pm J_1/2$ result in rotations about orthogonal axes, $\sigma_{Z}^{(1)}\sigma_{Z}^{(2)} \pm  \sigma_{X}^{(1)}$. Thus we can generate any rotation within the SU(2) subgroup by a piecewise constant pulse of three segments, again making use of Euler decomposition. In particular, to produce a $\pm\frac{\pi}{2}$ rotation of qubit 1 about $x$, we simply need to find the $t_i$ such that
 \begin{multline}\label{eq:X1}
     e^{-i\frac{J_1 t_1}{4}\left(\sigma_{Z}^{(1)}\sigma_{Z}^{(2)} +  \sigma_{X}^{(1)}\right)}e^{-i\frac{J_1 t_2}{4}\left(\sigma_{Z}^{(1)}\sigma_{Z}^{(2)} -  \sigma_{X}^{(1)}\right)}\\
     \times e^{-i\frac{J_1 t_3}{4}\left(\sigma_{Z}^{(1)}\sigma_{Z}^{(2)} +  \sigma_{X}^{(1)}\right)} 
     = e^{-i \frac{\pm\pi}{4}   \sigma_{X}^{(1)}}
 \end{multline}
There are multiple solutions, but the one with the minimum total elapsed time corresponds to a $\pi/2$ rotation, with time steps
\begin{equation}\label{eq:X1times}
    t_1=t_3 = \frac{\sqrt{2} \arccot{\sqrt{2}}}{J_1}; \hspace{2mm}
    t_2=\frac{5\sqrt{2}\pi}{3 J_1}.
\end{equation}

Unless one of the $J_i$ is somewhat tunable, the total elapsed time, $t_1+t_2+t_3$, will generally not be an integer multiple of $2\pi/J_2$. Because of this we must add a fourth segment to the evolution to avoid entangling qubits 2 and 3, but the effect of the fourth stage must be an identity in the SU(2) subgroup so as not to ruin the $\pi/2$ rotation.  This can be accomplished by setting the driving amplitude to
 \begin{equation}
     \Omega= 2 \sqrt{ \left( \frac{m\pi}{t_4}\right)^2-\left( \frac{J_1}{4}\right) ^2}
 \end{equation}
for a time 
\begin{equation}
     t_4 = \frac{2 n \pi}{J_2}-t_1-t_2-t_3
\end{equation} 
where $m$ and $n$ are integers to be chosen so that $t_4$ is as small as possible while still being positive (and also respecting any bounds on the maximum amplitude of $\Omega$). Combining the four evolution segments above produces a $\pi/2$ rotation about $X$ on qubit 1 and an $n\pi$ rotation about $Z$ on qubits 2 and 3.  For odd $n$, those extra $z$ rotations on qubits 2 and 3 can be instantaneously compensated by virtual $z$ rotations. 

The total gate time is thus $2 m\pi/J_2$ and the largest driving amplitude used is 
\begin{equation}\label{eq:Omegamax}
    \Omega_{\text{max}} = \max{\left(\frac{J_1}{2}, 2\sqrt{\frac{n^2\pi^2}{\left(\frac{2m\pi}{J_2}-\frac{\eta}{J_1}\right)^2}-\left(\frac{J_1}{4}\right)^2}\right)},
\end{equation}
where $\eta = 2\sqrt{2}\arccot{\sqrt{2}}+5\sqrt{2}\pi/3 \approx 9.14$.
In the realistic case where the gate time is constrained by the available ESR power rather than the exchange coupling and $J_1\approx J_2 \approx J$, if one maximizes the speed by fixing the exchange to be $J=2\Omega_{\text{max}}$, the shortest gate that respects the ESR constraint is obtained with $m=3, n=1$, yielding total gate time $3\pi/\Omega_{\text{max}}$. However, in the next subsection, in order to efficiently rotate the middle qubit, we will assume that $\Omega$ can be as large as $J$, so to keep a fixed value of exchange for all rotation types it is better to choose $J= \Omega_{\text{max}}$, in which case the total gate time doubles to $6\pi/\Omega_{\text{max}}$.

A rotation about $   \sigma_{X}^{(3)}$ can obviously be produced by driving resonantly with qubit 3 following the same procedure as above.

\subsection{\protect{$\frac{\pi}{2}$ $x$} Rotation of Center Qubit}
It is also possible to do a local rotation of qubit 2 by single-tone driving at its resonant frequency, $\omega_1 = E_z^{(2)}$, such that $\Omega_1^{(2)}=\Omega$ and $\Omega_2^{(i)}=0$. Here we rewrite the Hamiltonian as a sum of four mutually commuting terms, each belonging to a separate $\mathfrak{su}(2)$ subalgebra as previously noted \cite{Gullans_2019},
\begin{equation}
    H = H_{++}+H_{+-}+H_{-+}+H_{--}
\end{equation}
where
\begin{align}
     H_{++} &= \frac{J_1 + J_2}{8}Z_{++}+\frac{\Omega}{4} X_{++} \label{eq:Hpp}\\
     H_{+-} &= \frac{J_1 - J_2}{8}Z_{+-}-\frac{\Omega}{4} X_{+-} \label{eq:Hpm}\\
     H_{-+} &= \frac{-J_1 + J_2}{8}Z_{-+} -\frac{\Omega}{4} X_{-+} \label{eq:Hmp}\\
     H_{--} &= -\frac{J_1+J_2}{8}Z_{--}+\frac{\Omega}{4} X_{--} \label{eq:Hmm}
\end{align}
and
\begin{align}
    Z_{s_1 s_2} &=\frac{1}{2} \left(\sigma_{Z}^{(1)}+s_1 I\right) \sigma_{Z}^{(2)}  \left(\sigma_{Z}^{(3)}+s_2 I\right) \\
    X_{s_1 s_2} &= \frac{1}{2}\left(\sigma_{Z}^{(1)}+s_1 I\right) \sigma_{X}^{(2)}  \left(\sigma_{Z}^{(3)}+s_2 I\right),
\end{align}
where $s_i \in \{+,-\}$. The Overhauser fields for the first and third qubits can be written in terms of the $\mathfrak{u}(1)$ generators $Q_{s_1 s_2} = \frac{1}{2} \left(\sigma_{Z}^{(1)}+s_1 I\right) I  \left(\sigma_{Z}^{(3)}+s_2 I\right)$, which commute with $H$ and thus cannot be corrected. For simplicity, we consider the case where the couplings are essentially equal, $J_1=J_2=J$.  This may occur naturally under precise fabrication, but small deviations can also be tuned to zero by a small adjustment of the detuning or barrier voltages. The $+-$ and $-+$ subalgebras then reduce to $\mathfrak{u}(1)$s and the evolution in them is easily accounted for.

The $\sigma_{X}^{(2)}$ generator in terms of the above generators is
\begin{equation}
    \sigma_{X}^{(2)} = \frac{1}{2}\left(X_{++}-X_{+-}-X_{-+}+X_{--}\right),
\end{equation}
so the total desired evolution can be decomposed as
\begin{equation}
     e^{-i \frac{\pi}{4} \sigma_{X}^{(2)}} = e^{-i \frac{\pi}{8} X_{++}} e^{-i (-\frac{\pi}{8}) X_{+-}}  e^{-i (-\frac{\pi}{8}) X_{-+}}  e^{-i \frac{\pi}{8} X_{--}}.
\end{equation}
We begin by finding a pulse sequence that creates the desired $\pi/4$ $X_{++}$ rotation, toggling $\Omega$ between $\pm J$ and using the same type of Euler decomposition as in the previous subsection,
 \begin{multline}
     e^{-i \frac{J t_1}{4}\left(Z_{++}+X_{++}\right)}  e^{-i \frac{J t_2}{4}\left(Z_{++}-X_{++}\right)} \\
      e^{-i \frac{J t_3}{4}\left(Z_{++}+X_{++}\right)} = e^{-i\frac{\pi}{8}X_{++}}.
 \end{multline}
The solution with the minimum elapsed time is
 \begin{align}
     t_1=t_3&=\frac{\sqrt{2}\arctan{(1-1/\sqrt{2})}}{J}, \\ t_2&=\frac{2\sqrt{2} (\pi-\arctan{\sqrt{\frac{7-4\sqrt{2}}{17}}})}{J}
 \end{align}

Of course, this pulse sequence also produces a evolution in each of the other three subspaces at the same time. Fortunately, the evolution in the $--$ subspace produced is also the desired $\pi/4$ rotation due to the similarity of the Hamiltonians in the $--$ and $++$ subalgebras.  
The accompanying evolution in the $+-$ and $-+$ subspaces though is not generally equivalent to a $-\pi/4$ rotation. So, as in the previous subsection, we add one final step to the sequence such that the pulse area under $\Omega \left(t\right)$ for the entire sequence is $\frac{\pi}{4}$. Of course, this final step must also produce an identity in the $++$ and $--$ subspaces, so as not to ruin the $\pi/4$ rotations already produced there. These two conditions can be written as
 \begin{align}
     \frac{J}{4}(t_1-t_2+t_3) + \frac{\Omega}{4}t_4 &= \frac{\pi}{8} \\
     t_4 \sqrt{\left( \frac{J}{4}\right)^2+\left(\frac{\Omega}{4}\right)^2} &= \pi.
 \end{align}\par
The solution then for the length and drive amplitude of the final time step is
\begin{align}
     \Omega &= \frac{J(\pi-2 J \tau)}{\sqrt{\left(9\pi-2J\tau\right)\left(7\pi+2J \tau\right)}} \approx 0.998 J
     \\
     t_4 &= \frac{\sqrt{\left(9\pi-2J\tau\right)\left(7\pi+2J \tau\right)}}{2J} \approx 8.896/J
\end{align}
where $\tau = t_1-t_2+t_3$. Combining these four segments produces a $\pi/2$ rotation about $\sigma_{X}^{(2)}$ in a total time of about $18/J$.
 
 \subsection{{\sc cz} Gate on Nearest Neighbors}\label{subsec:uncorrectedCZ}
Performing a {\sc cz} gate on neighboring qubits, e.g., qubits 2 and 3, is considerably simpler, only requiring resonant single-tone driving of qubit 1, $\omega_1 = E_z^{(1)}$, such that $\Omega_1^{(1)}=\Omega$ and $\Omega_2^{(i)}=0$.  Then the Hamiltonian is the same as Eq.~\eqref{eq:CZdecomp1}-\eqref{eq:CZdecomp2b}.
The terms in the $\mathfrak{su}(2)$ part can be chosen so as to produce an identity operation while the $\mathfrak{u}(1)$ part produces the desired entanglement. To be specific,
\begin{equation}
e^{-i\left(\frac{J_1}{4} \sigma_{Z}^{(1)}\sigma_{Z}^{(2)} + \frac{J_2}{4} \sigma_{Z}^{(2)}\sigma_{Z}^{(3)} +\frac{ \Omega}{2} \sigma_{X}^{(1)}\right)t} = e^{-i \left(2m+1\right)\frac{\pi}{4}\sigma_{Z}^{(2)}\sigma_{Z}^{(3)}}   
\end{equation}
when 
\begin{equation}
    t=\frac{(2m+1)\pi}{J_2}
\end{equation} 
and
\begin{equation}
    \Omega=2\sqrt{\left(\frac{n J_2}{2m+1}\right)^2-\left(\frac{J_1}{4}\right)^2}
    \label{eq:uncorrectedCZOmega}
\end{equation} 
for integer $m$ and $n$.
Note that for $J_1\approx J_2$, the ESR strength required can be kept less than $J/2$ by choosing $m=1, n=1$.

\subsection{Composite pulses for dynamical correction}
We briefly remark that the above are all uncorrected sequences that accomplish gates by taking advantage of time evolution in commuting subspaces without worrying about noise effects. If we include some error in exchange, $J\rightarrow J+\delta J$, this will clearly produce some error in the previous results. This error can be corrected for, to some arbitrary order in $\delta J$, with the direct application of {\sc supcode} pulse sequences \cite{Wang_2012,Kestner_2013,Wang_2014,Wang_2014PRB}
by individually replacing each piece of our pulse sequence with a corresponding corrected {\sc supcode} pulse. While straightforward, that would produce inefficiently long sequences, as {\sc supcode} is designed to do more than is actually necessary in this case, correcting for errors on both the coupling and the driving field simultaneously. Besides that, instead of correcting each piece of our Euler decompositions separately, we should instead correct the entire rotation with a single optimal application of {\sc supcode}. In a future investigation, we will explore the degree to which the protocol can be optimized for this particular case, resulting in more efficient correction. However, for the particular case of a {\sc cz} gate we explicitly show below how pulse shaping can be used to perform dynamical error correction by using the SU(2) dynamics generated by Eq.~\eqref{eq:CZdecomp2b} rather than the simpler U(1) dynamics generated by Eq.~\eqref{eq:CZdecomp2a} that we used above.

\section{Pulse shaping for error correction} \label{sec:pulsetheory}
In this section we will describe the procedure for creating a {\sc cz} gate that is robust against errors using two-tone driving. Both exchange noise, $\delta J$, and multiplicative pulse amplitude error, $\delta \Omega(t) = \Omega(t) \delta \alpha$, can be corrected via a formalism introduced by Barnes et al.~\cite{Barnes2015}. We briefly summarize the relevant results for completeness below, following the presentation of Ref.~\cite{Gungordu2020}, before moving to the specific application of a {\sc cz} in a three-qubit system.

\subsection{Background}\label{subsec:background}
The formalism of Ref.~\cite{Barnes2015} applies to any $\mathfrak{su}(2)$ Hamiltonian with a constant term on one generator and a controllable term on another, including a noisy analog of Eq.~\eqref{eq:CZdecomp2b} with multiplicative driving amplitude noise and exchange noise,
\begin{equation}
    H = \frac{\Omega(t)}{2}\left[1 + \delta \alpha\right] \sigma_{X} + \frac{J+\delta J}{4} \sigma_{Z}.
\end{equation}
The time dependence of the system is then re-parameterized in terms of a new variable $\chi$ which is related to time via
\begin{equation}
\frac{J}{4}t = \hbar \int_0^{\chi_f} d\chi \sqrt{1+(\Phi'(\chi)\sin(2\chi))^2}
    \label{eq:timechi}
\end{equation}
Here $\Phi(\chi)$ is a function which is free to choose within some constraints which follow from  Ref.~\cite{Barnes2015}. Choosing $\Phi(\chi)$ determines the Hamiltonian and therefore the evolution operator $U$. One constraint on $\Phi(\chi)$ is that $\Phi(0)=\Phi'(0)=0$, which ensures that $U$ is an identity for zero elapsed time.
The pulse shape $\Omega(t)$ can be recovered from the chosen function  $\Phi(\chi)$ in terms of $\Omega(t)=\tilde{\Omega}\left(\chi(t)\right)$ using Eq.~\eqref{eq:timechi} and
\begin{multline}
     \tilde{\Omega}(\chi)=-\frac{J}{2} \sin(2 \chi) 
     \\
     \times \frac{\Phi''(\chi)+4\Phi'(\chi)\cot(2\chi)+2(\Phi'(\chi))^3\sin(4 \chi) }{2(1+(\Phi'(\chi)\sin(2\chi))^2)^{3/2}}
\end{multline}
The conditions for canceling the exchange noise, $\delta J$, to first order are
\begin{equation}
\begin{split}
    \sin(4\chi_f)+8 e^{-2 i \Phi(\chi f)} \int_0^{\chi_f} d\chi \sin^2(2 \chi)e^{2 i \Phi(\chi )}=0\\
    \int_0^{\chi_f} d\chi \sin^2(2 \chi) \Phi'(\chi)=0
    \end{split}
    \label{eq:deltaJnoisecondition}
\end{equation}
and the conditions for canceling the driving noise, $\delta \alpha$, to first order are
\begin{equation}
    \begin{split}
     \int_0^{\chi_f} d\chi \sin(2 \chi) \Tilde{\Omega}(\chi) e^{2 i \Phi(\chi )} \sqrt{1+(\Phi'(\chi)\sin(2\chi))^2}=0\\
     \int_0^{\chi_f} d\chi \cos(2\chi) \Tilde{\Omega}(\chi)  \sqrt{1+(\Phi'(\chi)\sin(2\chi))^2}=0.
    \end{split}
    \label{eq:deltaEnoisecondition}
\end{equation}
The second condition in Eq.~\eqref{eq:deltaJnoisecondition} and Eq.~\eqref{eq:deltaEnoisecondition} can be satisfied by simply choosing an odd function for $\Phi(\chi)$, which implies $\Omega(-t)=-\Omega(t)$, and extending the pulse duration to be from $-t_f$ to $t_f$. In this case the evolution from $-t_f$ to $t_f$  becomes
\begin{equation}
\label{eq:Ufthetaphi}
     U(t_f;-t_f)=e^{- i \frac{\theta}{2}(\cos(\phi)\sigma_{Z}+\sin(\phi)\sigma_{Y})},
\end{equation}
where
\begin{align}
        \phi&= \sign(\Phi'(\chi_f))\arcsec(\sqrt{1+(\Phi'(\chi_f)\sin(2\chi_f))^2})\\
        \theta&=4\chi_f.
    \label{eq:decomp}
\end{align}
Following Ref.~\cite{Gungordu2020}, in this paper we will choose the ansatz
\begin{equation}\label{eq:ansatz}
    \Phi(\chi) = a_1 \chi^2+\sign(\chi) \left[a_2\chi^3 +\sum_{n=1}^{8} b_n \sin(\frac{ n \pi \chi}{\chi_f})\right].
\end{equation}
This ansatz automatically satisfies the initial condition $\Phi(0)=\Phi'(0)=0$. To ensure that $\Omega(t)$ vanishes at $t=0$ and $t=\pm t_f$, $a_1$ and $a_2$ must be
\begin{align}
        a_1 &=\frac{\tan \phi}{2 \sin(2\chi_f)}(1+\chi_f\cot(2\chi_f)(1+\sec^2\phi))\\
        a_2 &=-\frac{\tan \phi}{3 \chi_f^2  \sin(2\chi_f)}(1+2 \chi_f\cot(2\chi_f)(1+\sec^2\phi))
\end{align}
except for the case $\theta=2 n \pi$, i.e., $\chi_f = n \pi/2$, where $n\in \mathbb{Z}$, which requires $a_1=a_2=0$ for any $\phi$.
The robustness conditions Eq.~\eqref{eq:deltaJnoisecondition}-\eqref{eq:deltaEnoisecondition} are then satisfied by optimizing the free parameters, $b_n$, in Eq.~\eqref{eq:ansatz}.

\subsection{Corrected {\sc cz} Gate on Nearest Neighbors}\label{sec:pulseresult}
Although the formalism summarized above could directly be used to correct noise within the $\mathfrak{su}(2)$ part of the Hamiltonian \eqref{eq:CZdecomp2b}, there is also the $\mathfrak{u}(1)$ part \eqref{eq:CZdecomp2a} in which no dynamical correction of the exchange noise $\delta J_2$ is possible. Thus, to create a truly robust {\sc cz} gate requires two-tone driving such that both of the outer qubits are addressed such that $\omega_1 = E_z^{(1)}$, $\Omega_1^{(1)}=\Omega_1$, $\omega_2 = E_z^{(3)}$, and $\Omega_2^{(3)}=\Omega_2$. The Hamiltonian then contains two commuting $\mathfrak{su}\left(2\right)$ parts, $H=H_{\mathfrak{su(2)}}^1 + H_{\mathfrak{su(2)}}^2$, where
\begin{align}\label{eq:erroshamiltonian1}
H_{\mathfrak{su(2)}}^1= \frac{J_1+\delta J_1}{4} \sigma_{Z}^{(1)}\sigma_{Z}^{(2)} +\frac{\Omega_1}{2}(1+\delta \alpha_1) \sigma_{X}^{(1)} \\
\label{eq:erroshamiltonian2}
H_{\mathfrak{su(2)}}^2= \frac{J_2+\delta J_2}{4} \sigma_{Z}^{(2)}\sigma_{Z}^{(3)} + \frac{\Omega_2}{2}(1+\delta \alpha_2) \sigma_{X}^{(3)}.
\end{align}
Then the formalism described in Sec.~\ref{subsec:background} must be adjusted to ensure that the two copies of SU(2) are simultaneously corrected \emph{with the same pulse time}, $2t_f$.

The final evolution in both $SU(2)$s is like that given in Eq.~\eqref{eq:Ufthetaphi}. To create a {\sc cz} gate between the first and second qubit the angles for $H_{\mathfrak{su(2)}}^1$ are required to be 
\begin{equation}\label{eq:angles1}
    \theta_1=(k_1 +1/2)\pi, \qquad \phi_1 = 0,
\end{equation} 
while the angles for $H_{\mathfrak{su(2)}}^2$ are required to be 
\begin{equation}\label{eq:angles2}
    \theta_2= k_2 \pi, \qquad \phi_2=0,
\end{equation} 
where $k_i$ is any positive integer. The angle $\phi_2=0$ is chosen in this case since, by Eq.~\eqref{eq:Ufthetaphi}, if $k_2$ is even, $\phi_2$ is arbitrary, and if $k_2$ is odd then $\phi_2=0$ corresponds to an extra local $e^{i \frac{\pi}{2}  \sigma_{Z}^{(2)}\sigma_{Z}^{(3)}}=\sigma_{Z}^{(2)}\sigma_{Z}^{(3)}$ rotation which can be compensated with virtual $z$ rotations. Finally, note that one could similarly target $x$ rotations on the outer qubits within this formalism, but not rotations of the middle qubit. For that reason, we restrict our attention to the {\sc cz} gate in this work, and leave generic robust local gates as a topic for future investigation.

Dynamical correction of all error terms in Eqs.~\eqref{eq:erroshamiltonian1} and \eqref{eq:erroshamiltonian2} means that pulse shapes for $\Omega_1(t)$ and $\Omega_2(t)$ must both satisfy (or at least nearly satisfy) the conditions in Eqs.~\eqref{eq:deltaJnoisecondition} and \eqref{eq:deltaEnoisecondition}.
The optimizations of the two sets of free parameters, $\{b_n^{(1)}, b_n^{(2)}\}$, must respect the constraint that the pulses need to be of the same time length.
Equation \eqref{eq:timechi} determines the time length, which depends on the optimized pulse shape as well as $J_1$ and $J_2$. 
So to be able to match length we must know the ratio of $J_1$ and $J_2$. 
For the results we present we have assumed $J_1/J_2 \approx 1$, though this value is not necessary, and by construction our results will be robust against slight variations in the ratio. Also note that since the pulse times are parameterized in units of $1/J_i$ in Eq.~\eqref{eq:timechi}, a slight difference in the pulse lengths of $\Omega_1$ and $\Omega_2$ acts in the same way as $\delta J_i$, and the pulse is again robust by construction to such perturbations, so the pulse lengths do not need to be exactly the same, only close.

The optimized coefficients for this pulse are
\begin{align}
    b^{(1)}&\approx\{1.3, -0.71, -0.6, -0.18, -0.13, -0.07, -0.06, 0.03\}\\
    b^{(2)}&\approx\{3.0, -0.26, 0.6, 0.34, -0.07, -0.13, -0.02, -0.07\}
    \label{eq:bboth}
\end{align}
\begin{figure}[tp!]
    \centering
    \includegraphics[width=\linewidth]{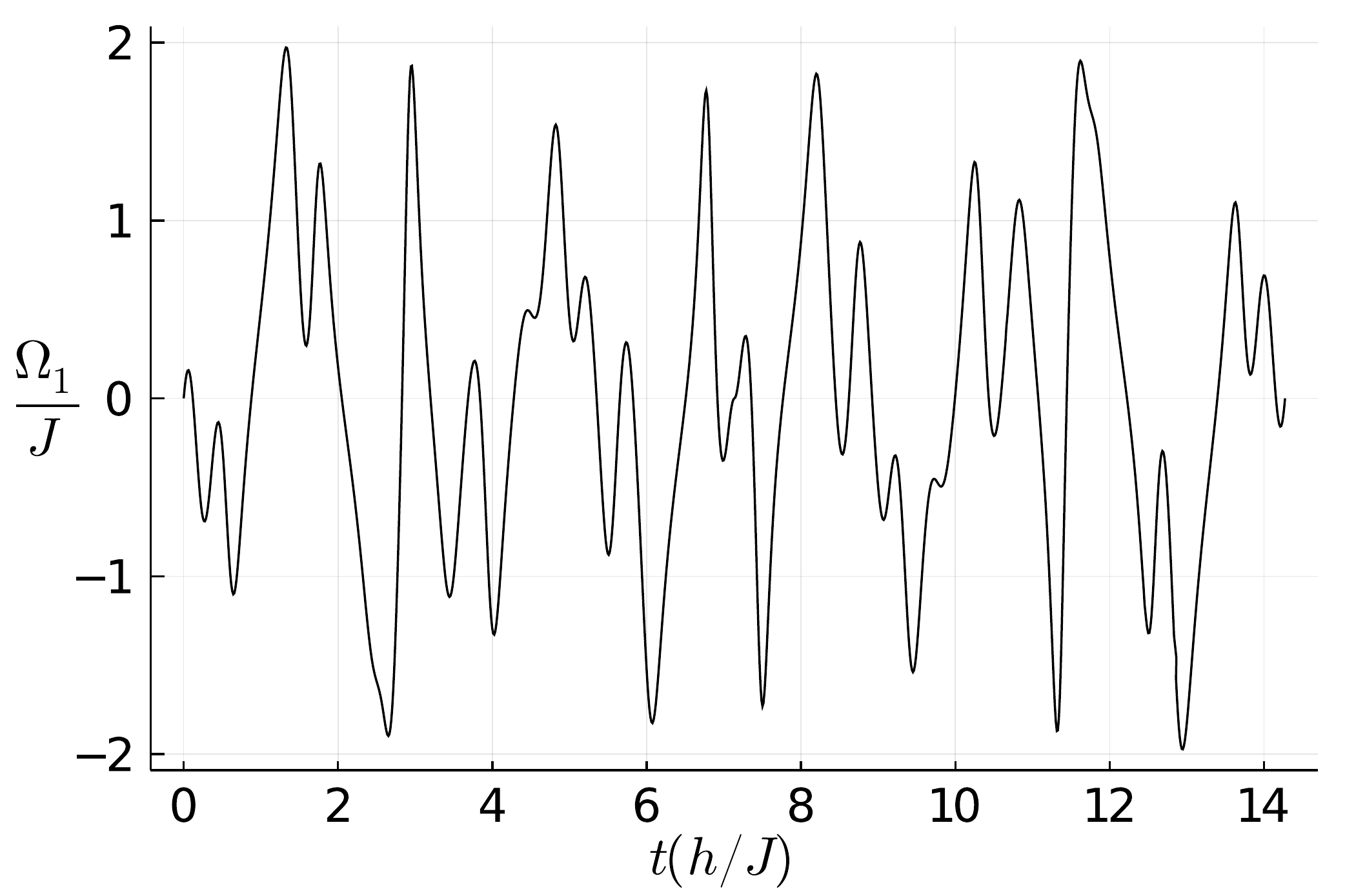}
    \includegraphics[width=\linewidth]{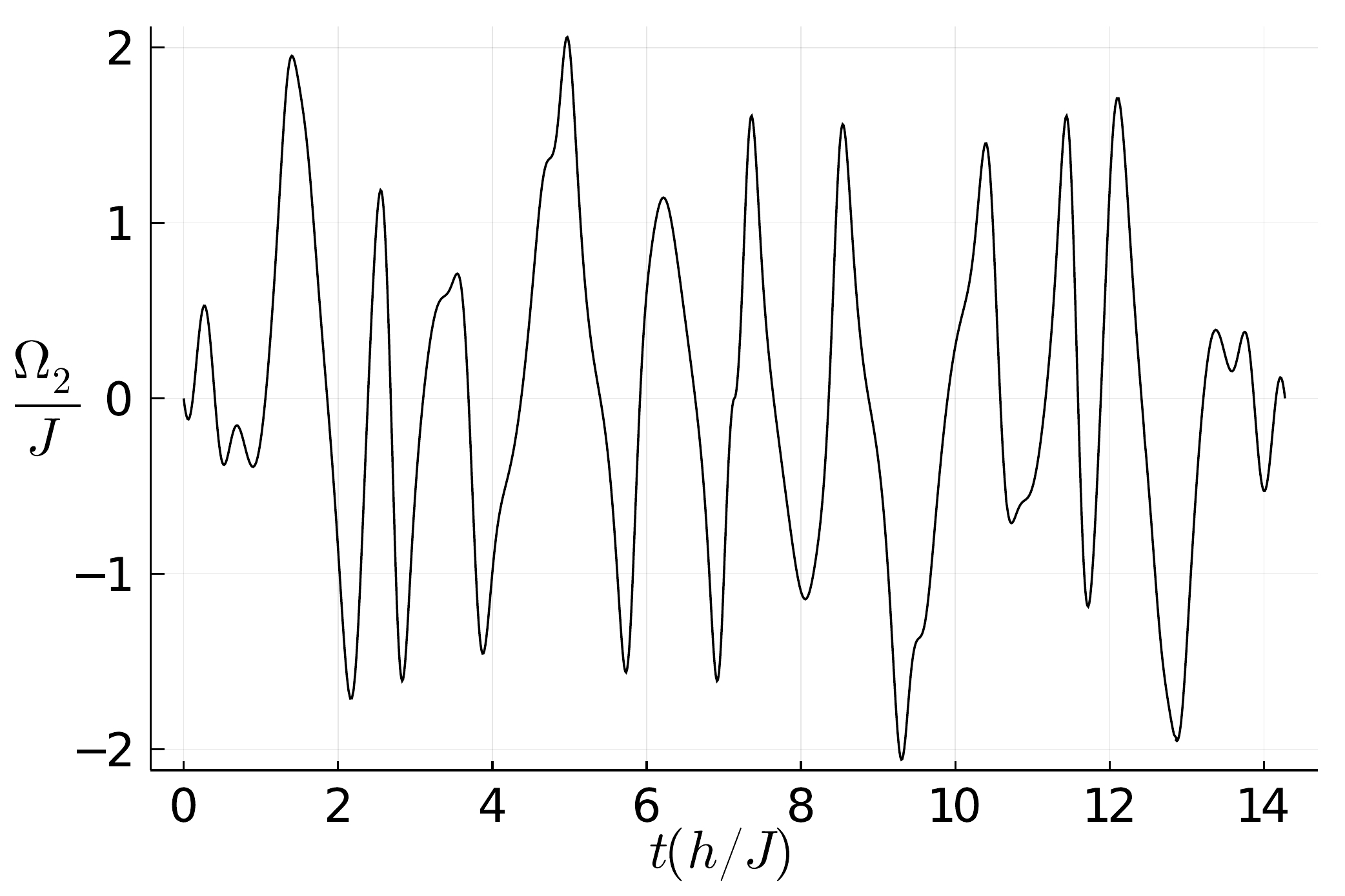}
    \caption{Pulse shapes $\Omega(t)/J$ versus time in units of $h/J$. Top: Amplitude of tone resonant with qubit 1, resulting in a $\pi/4$ rotation around $\sigma_{Z}^{(1)}\sigma_{Z}^{(2)}$. Bottom: Amplitude of tone resonant with qubit 3,  which results in an identity in the $\sigma_{Z}^{(2)}\sigma_{Z}^{(3)}+\sigma_{X}^{(3)}$ SU(2).}
    \label{fig:CZbothtime}
\end{figure}
where $b^{(1)}$ are the coefficients for $\Omega_1(t)$ and $b^{(2)}$ for are the coefficients for $\Omega_2(t)$. These coefficients were found by using the randomly-seeded non-linear gradient-free global optimizer ISRES from the NLopt package in Julia. We enforced constraints $\Omega_i/J_i<1$ and $2t_f J_i < 15$. Manually adjusting these constraints as optimization hyper-parameters, we have found that these constraints produce the fastest pulses in the case where the pulse length is limited by the physically attainable ESR strength. For instance, if one doubles the maximum allowed value of $\Omega_i/J_i$ to 2, one can find a solution with somewhat smaller values of $2t_f J_i$, but if one cannot increase $\Omega_i$ beyond its prior value, that doubling of the ratio must come from by halving $J_i$, resulting in a net increase of the pulse time $2t_f$. Conversely, reducing the maximum value of the $\Omega_i/J_i$ allows larger $J_i$ but requires much larger values of $2 t_f J_i$. We have empirically determined that the numerical choices in the constraints above give roughly optimal results.

For $\Omega_2$ no solutions were found for values of $k_2<12$ when requiring the integrals in  Eqs.~\eqref{eq:deltaJnoisecondition} and \eqref{eq:deltaEnoisecondition} to sum to less than $10^{-3}$ after about 80 million iterations of the optimization algorithm. 
Since the pulse length generally scales with $k_i$ in Eqs.~\eqref{eq:angles1}-\eqref{eq:angles2}, we chose the smallest values for which a good solution could be found, $k_1=k_2=12$.
The times of these two pulse shapes are not exactly equal, but the difference is small; they would match exactly for $J_1=1.01 J_2$, and the error introduced by the mismatch is negligible.
The corresponding pulse shapes are shown in Fig.~\ref{fig:CZbothtime}. Note that it requires ESR amplitudes as large as the exchange, so if we limit the exchange to $J=h\times 2$MHz so that we do not require more than $h\times 2$MHz on either ESR tone, the total pulse time is about 7$\mu$s. This isabout 28 times longer than the uncorrected pulse time of $h /(2 J)= 0.25\mu s$, and so one may worry about the prolonged interaction with the environment.However, with observed $T_1$ decoherence times in silicon spin qubits of, e.g., 3s \cite{Simmons2011}, relaxation is not a limiting factor and the corrected pulse is expected to offer orders of magnitude reduction in infidelity. We can show this rigorously by doing a master equation analysis which describes the dynamics of the density matrix $\rho$ as 
\begin{equation}
    \frac{d \rho}{dt} = - i \left[ H,\rho \right] + \sum_{j}^3 \frac{1}{T_1} D[L_j]\rho
\end{equation}
where $L_j$ is the lowering operator for the $j$-th qubit and $D$ is the damping superoperator $D[A]\rho=2 A \rho A^\dagger-\frac{1}{2}A^\dagger A \rho -\frac{1}{2}\rho A^\dagger A$. The average fidelity can then be calculated using the method from Ref.~\cite{CABRERA2007}. Doing this analysis shows that the corrected pulse maintains a $10^{-4}$ average infidelity for $T_1$ as low as 500ms, well below experimentally achievable $T_1$ times.\par
\begin{figure}[tp!]
    \centering
    \includegraphics[width=\linewidth]{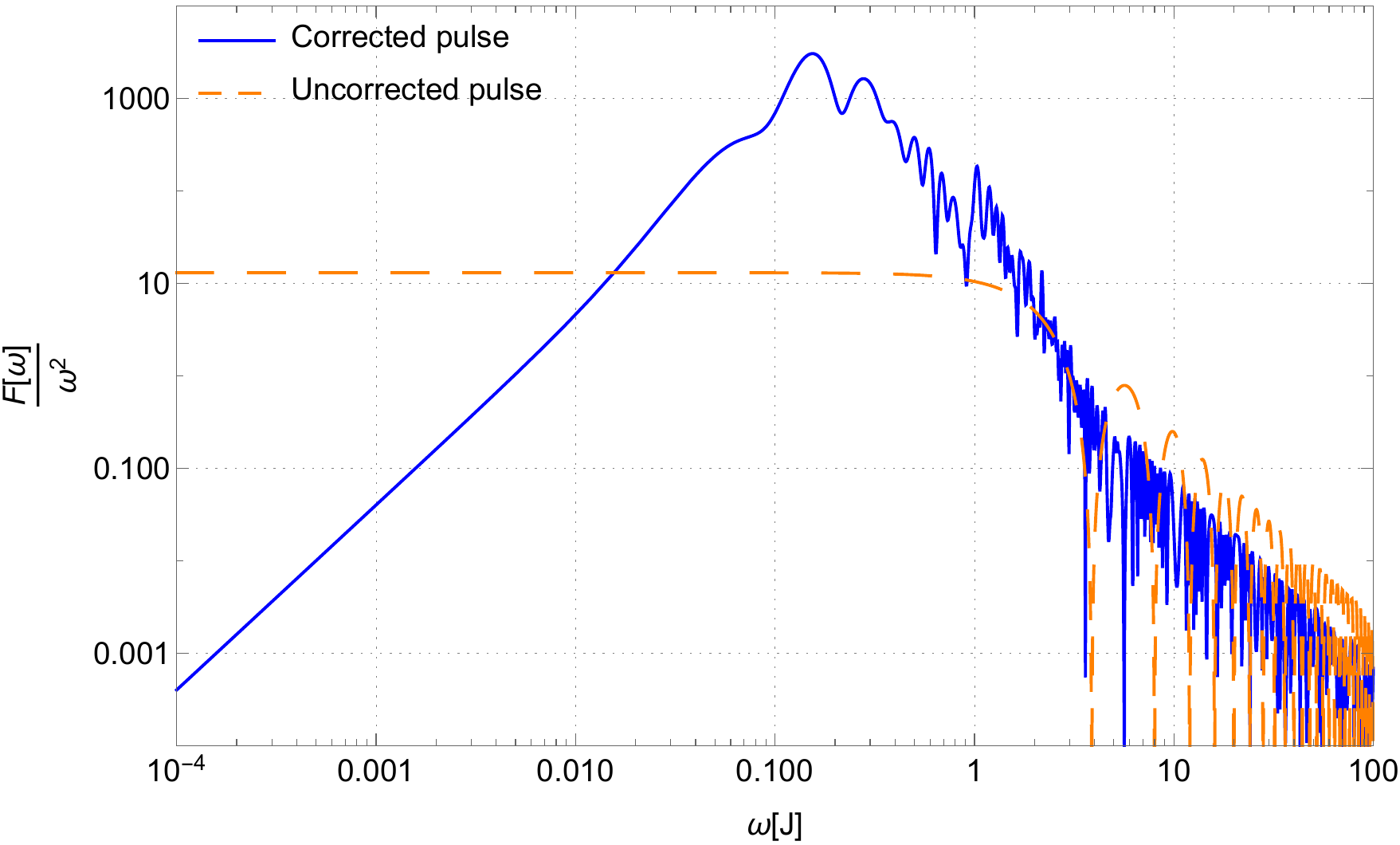}
    \caption{Filter function of the corrected pulse and uncorrected pulse vs the frequency $\omega$ in units of $J$.}
    \label{fig:filterfunction}
\end{figure}

The decoherence from $T_2$ processes is harder to capture since the noise is non-Markovian and the master equation approach cannot be used. Therefore the filter function of the pulse, $\text{F}(\omega)$ as defined in Ref.~\cite{Green_2013}, was calculated and plotted in Fig.~\ref{fig:filterfunction} in order to take into account the typical 1/$f$ power spectral density (PSD) of charge noise. The average infidelity over noise realizations, $\mathcal{F}_{\text{av}}$, was calculated by integrating the filter function multiplied by a noise PSD of $A_0^2/\omega$ with an infrared frequency cutoff, $\omega_{ir}$,
\begin{equation}
    \mathcal{F}_{av}\approx1-\frac{1}{2\pi}\int^{\infty}_{\omega_{ir}} \frac{A_0^2}{\omega}\frac{\text{F}(\omega)}{\omega^2} 
    \text{d}\omega
\end{equation}
The strength of the noise PSD, $A_0$, was determined by using its relation to the Carr-Purcell-Meiboom-Gill (CPMG) decoherence time, $\frac{0.85 (A_0)^2}{2 \pi n}\approx
\frac{1}{(T_{2}^{\text{CPMG}})^2}$ where $n$ is the number of $\pi$-pulses used to measure $T_2^{\text{CPMG}}$ \cite{Cywinski2008}. The resulting infidelity is $5 \times 10^{-4}$ choosing $A_0$ such that $T_2^{\text{CPMG}}=28$ms for a CPMG series of 500 $\pi$-pulses as measured in Ref.~\cite{Veldhorst_2014} and an infrared frequency cutoff of $10^{-5}$Hz corresponding to daily calibration. (As quantum devices get larger and more complicated, calibrations become more time-consuming and less likely to be done frequently.) This is an improvement over an uncorrected pulse which has an infidelity of $1.2 \times 10^{-3}$ for this $T_2$ and infrared cutoff. Thus, despite the much longer duration of the corrected pulse, it is still worth doing. Furthermore, the performance of the corrected pulse is also significantly better than the estimate above if the noise is more heavily weighted at low frequencies than 1/$f$, as observed, e.g., in Ref.~\cite{Struck_2020}, or if the calibration is imperfect, effectively resulting in additional quasi-static noise.

\begin{figure}[tp!]
    \centering
    \includegraphics[width=\linewidth]{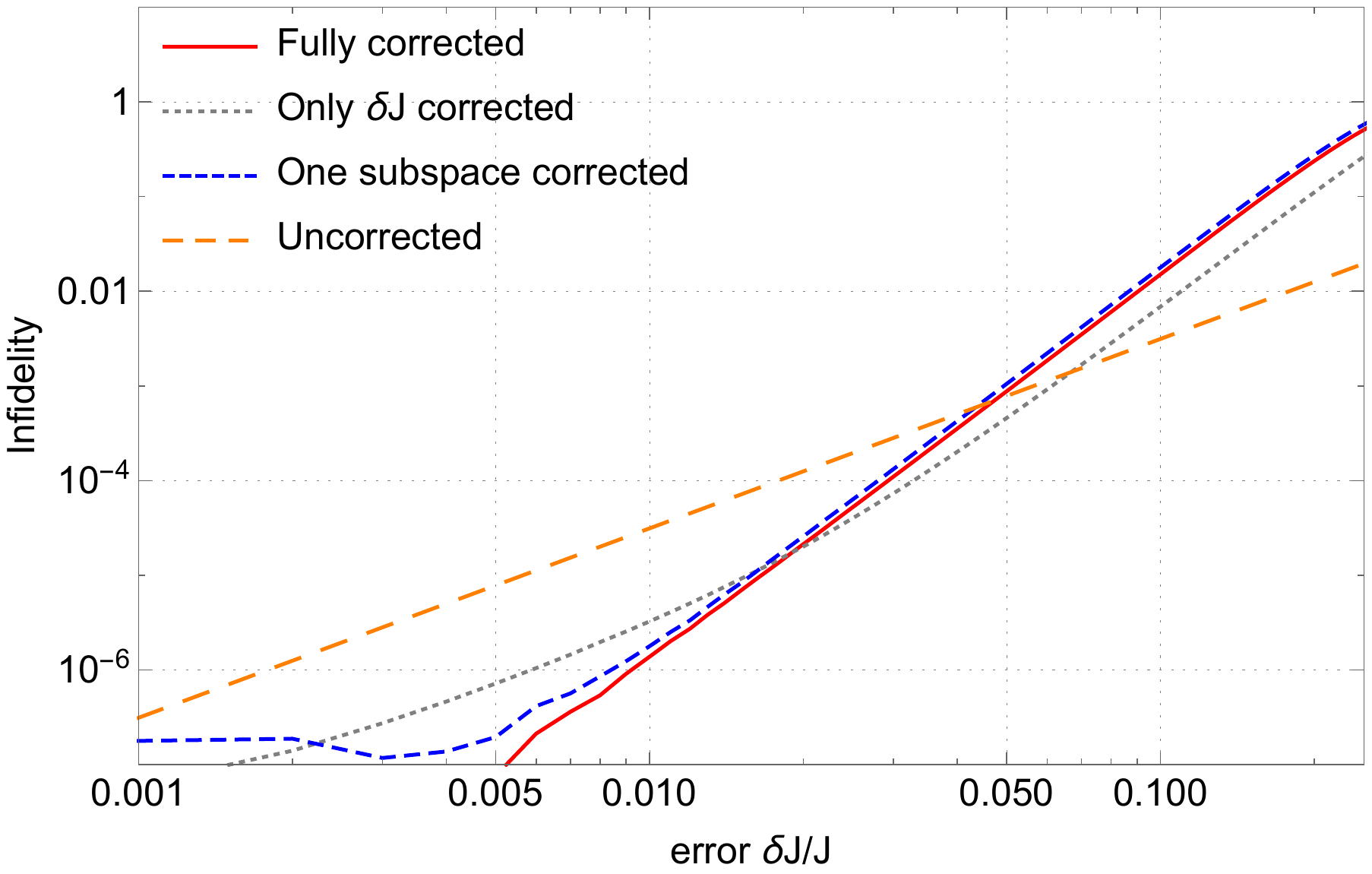}
    \includegraphics[width=\linewidth]{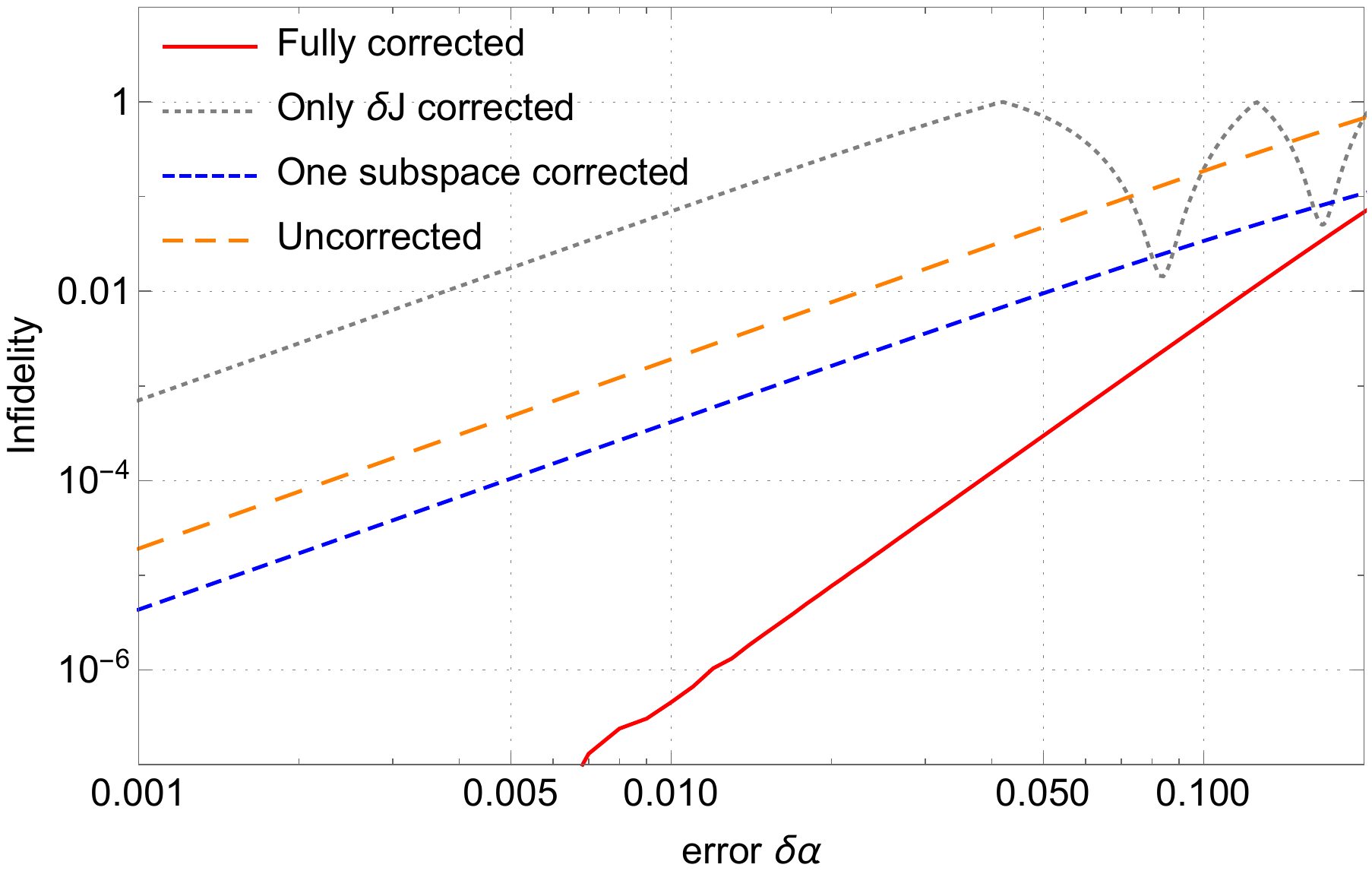}
    \caption{Infidelity of the error corrected {\sc cz} and the uncorrected {\sc cz} vs quasi-static exchange noise $\delta J$ (top) and driving noise $\delta \alpha$ (bottom).}
    \label{fig:CZbotherror}
\end{figure}

We have also plotted infidelity vs a quasi-static noise strength in Fig.~\ref{fig:CZbotherror}. The infidelity plotted in Fig.~\ref{fig:CZbotherror} is defined as $1-\frac{|Tr(UU_t^{\dagger})|}{Tr(U_t U_t^{\dagger})}$ where $U$ is the actual noisy gate and $U_t$ is the desired target gate. The infidelity depends on the values of $\delta J_1$, $\delta J_2$, $\delta \alpha_1$, and $\delta \alpha_2$, but we have simplified the display by plotting for $\delta J_1=\delta J_2=\delta J$ and $\delta \alpha_1=\delta \alpha_2 = \delta \alpha$.
It is clear that this error-corrected pulse performs better than the uncorrected {\sc cz} pulse from Sec.~\ref{subsec:uncorrectedCZ} where $m=0$ and $n=1$ is chosen in Eq. \ref{eq:uncorrectedCZOmega} and $J_1=J_2=h\times2$MHz. Increasing the $J_i$ in the uncorrected pulse does not significantly change the infidelity as long as $\Omega$ is still limited to $h\times4$MHz for single-tone driving.

We also looked at alternative optimization objectives. For instance, one can correct the {\sc cz} gate against only exchange coupling noise, enforcing only Eq.~\eqref{eq:deltaJnoisecondition}. This yields similar performance for exchange coupling noise however becomes highly sensitive to amplitude error.
But, if the amplitude error is negligible, correcting only exchange noise requires less bandwidth and a slightly shorter pulse length of $\approx 6 \mu$s. Another approach for correcting just exchange noise is to note that the evolution produced by $H_{\mathfrak{su(2)}}^2$ is supposed to be an identity.
A simple pulse of strength $\Omega_2=2\sqrt{(\frac{2 \pi  k}{T/h})^2 -J_2^2/16}$ for any time $T$ and $k\in \mathbb{Z}$ results in an uncorrected identity with an infidelity that scales as $c_2 \delta J_2^2$ compared to $\sim \delta J_2^4$ of a robust shaped pulse, but with the very small coefficient $c_2= \frac{J_2^2 T^4}{2048 h^4 k^2 \pi^2}$. Taking $J_2=h\times2$MHz and $T=6.1\mu s$, $k=2$ is the maximum value for which the ESR strength $\Omega$ remains below $h\times2$MHz, and leads to $c_2$ being about 1/100 of the value one would get in the absence of driving letting $(J_2+\delta J_2)/4$ $\sigma_{Z}^{(1)}\sigma_{Z}^{(2)}$ evolve on its own.

This means that one may be able to get away with not doing any correction in that subspace, shaping only the $H_{\mathfrak{su(2)}}^1$ pulse without worrying about making pulse lengths match. This has the benefit of allowing for slightly shorter pulse lengths since searching for matching pulse length results in longer pulses. As shown in Fig.~\ref{fig:CZbotherror}, this approach works fairly well against exchange noise, though there is substantial performance degradation if amplitude error is appreciable.

A direct comparison of our corrected pulse to methods other than that of Sec.~\ref{subsec:uncorrectedCZ} is not possible simply because of the lack of other robust {\sc cz} methods for always-on exchange in a silicon three-qubit system. Furthermore, it is not known what the fundamental speed limit is in this case or how close we are to it given the control constraints and the robustness requirement. Quantum speed limits are usually considered for nonrobust gates, though some recent work has also found the minimum times for robust gates in a certain single-qubit scenario \cite{Zeng_2018}, in which case the minimal time was about twice as long as the corresponding nonrobust rotation. It is not unreasonable that our pulse is longer than the nonrobust one by a factor of 28 considering the complexity of the three-qubit control landscape and the more complicated form of errors that are being corrected, but we certainly do not rule out the possibility of a faster solution.

However, for the sake of context, note that an uncorrected three-qubit i-Toffoli gate has been considered in such a system \cite{Gullans_2019} with a gate time of about $2h/J$, i.e., roughly 7 times faster than our corrected three-qubit {\sc cz} pulse, but even in the absence of noise it is already an approximation (albeit quite a good approximation, with an average error of either 0.6\% or 0.03\%, depending on the specific parameters) and it is not robust to amplitude or exchange noise/miscalibrations. Alternatively, comparing to a robust {\sc cz}-gate in a two-qubit system from the same method we have generalized here \cite{Gungordu2020}, the pulse length in the two-qubit case is only slightly less at $13h/J$, and the infidelity scales the same vs quasi-static error as our three-qubit pulse but is better by a constant factor of about two. This is because two independent exchange errors, $\delta J_i$, and amplitude errors $\delta \alpha_i$ have to be corrected at once in our three-qubit pulse while only one of each error exist for the two-qubit pulse. Another robust {\sc cnot} gate in an exchange-coupled two-qubit system is the {\sc supcode} implementation for amplitude error and exchange error \cite{Kestner_2013} which achieves an improvement of infidelity of more than two orders of magnitude over an uncorrected gate, similar to the present improvements in Fig.~\ref{fig:CZbotherror}. However the {\sc cnot} gate in that method was much longer, taking about 300 times longer than the uncorrected gate, so our current method is much more efficient. Experimentally, robust pulses for a single qubit device using {\sc grape} have been implemented and achieved infidelities as low as $4\times 10^{-4}$\cite{Yang2019} with a pulse time of $8\mu$s compared to an uncorrected $2\mu$s square pulse. In comparison, our pulse incurs a higher time cost relative to the uncorrected pulse, but performs a much more complicated gate compared to the single-qubit case. Also note that our pulse time estimate of $7\mu$s with current ESR strengths is not experimentally forbidding, and is comparable to the pulse times already used with the more limited ESR Rabi frequency available in the device of Ref.~\cite{Yang2019}.

\section{Summary \& Conclusions} \label{sec:conclusions}
We have shown how to perform a full set of local gates and {\sc cz} gates in a linear array of three spin qubits with always-on exchange coupling. The {\sc cz} gate was dynamically corrected using shaped two-tone ESR.
In principle all local gates can also be corrected via existing {\sc supcode} pulse sequences, though that approach may be unwieldy in practice without further optimization. Single-qubit pulse shaping is expected to exhibit improvement similar to that of the {\sc cz} gate presented above, although the more complicated requirements of correcting 4 different copies of SU(2) simultaneously when local rotations of the middle qubit are needed poses a formidable challenge that will be a topic for future investigation.

\section*{Acknowledgements} \label{sec:acknowledgements}
    This research was sponsored by the Army Research Office (ARO), and was accomplished under Grant No. W911NF-17-1-0287. 

\bibliographystyle{apsrev4-1} 
\bibliography{refs}

\end{document}